\documentclass[preprintnumbers, floatfix, twocolumn,
preprintnumbers, PRD, superscriptaddress,nofootinbib]{revtex4-1}
\usepackage{CJK}
\usepackage{amsmath,amssymb,bm,mathrsfs}
\usepackage{mathtools,mathptmx,array,slashed}
\usepackage{eurosym}
\usepackage{amsfonts}
\usepackage{amsmath}
\usepackage{amssymb,epsf}
\usepackage{latexsym}
\usepackage{graphicx,epsfig}
\usepackage{amssymb}
\usepackage{subfigure}
\usepackage{epstopdf}
\usepackage[colorlinks=true,citecolor=blue,linkcolor=blue,urlcolor=black]{hyperref}
\usepackage{color}
\usepackage{float}

\graphicspath{{Images/}}

\renewcommand{\&}{\textup{\symbol{`\&}}}
\usepackage{hyperref}
\usepackage{amsfonts}
\usepackage{amsmath}

\usepackage{amssymb}
\usepackage{graphicx}%
\usepackage{bigints}
\usepackage{graphicx}
\usepackage{subfigure}
\usepackage{epsf}
\usepackage{bm}
\usepackage{amsmath}
\usepackage{amsfonts}
\usepackage{amssymb}
\usepackage{graphicx}
\usepackage{tabularx}
\usepackage{color}%
\providecommand{\U}[1]{\protect\rule{.1in}{.1in}}

\newcommand{\be}{\begin{equation}}
\newcommand{\ee}{\end{equation}}

\newcommand{\mincir}{\raise
-3.truept\hbox{\rlap{\hbox{$\sim$}}\raise4.truept\hbox{$<$}\ }}
\newcommand{\magcir}{\raise
-3.truept\hbox{\rlap{\hbox{$\sim$}}\raise4.truept\hbox{$>$}\ }}

\def\be{\begin{equation}}
\def\ee{\end{equation}}
\def\bea{\begin{eqnarray}}
\def\eea{\end{eqnarray}}
\def\ba{\begin{aligned}}
\def\ea{\end{aligned}}

\def\p{\partial}

\providecommand{\U}[1]{\protect\rule{.1in}{.1in}}

\usepackage{tikz,xcolor,hyperref}

\begin{document}
\title{Topological classes of thermodynamics of black holes in perfect fluid dark matter background}

\author{Muhammad Rizwan}
\email{mrizwan@numl.edu.pk}
\affiliation{Faculty of Engineering and Computer Sciences, National University of Modern
	Languages, H-9, Islamabad, Pakistan}
 \author{Kimet Jusufi}
\email{kimet.jusufi@unite.edu.mk}
\affiliation{Physics Department, State University of Tetovo, 
Ilinden Street nn, 1200, Tetovo, North Macedonia}

\begin{abstract}
In this paper we explore the topological classes of thermodynamics of a family of black holes. In particular we investigate the influence of distinct fields, including the electric field, non-linear magnetic field, along with the perfect fluid matter background that can mimic dark matter in large distances. In light of these considerations, we shall henceforth denote this fluid as perfect fluid dark matter (PFDM). Our analysis reveals that the winding and topological numbers for the Schwarzschild and Kerr black holes in PFDM background are the same as the Schwarzschild and Kerr black holes, however for the Kerr-AdS background in PFDM we obtain a different topological number compared to the Kerr black hole in PFDM.  Furthermore, we explore in details the interplay of electric charge and nonlinear magnetic charge, impacting the topological classes of thermodynamics both in the absence and presence of PFDM.  Interestingly, it is demonstrated that the topological numbers associated with the static Hayward black holes, both in the absence and presence of PFDM, deviate from those of the Schwarzschild black hole. This observation shows that the presence of a magnetic charge introduces an additional role and can alter the topological numbers. Finally, our study culminates with the comprehensive analysis of the topological numbers pertaining to the Hayward black hole, considering the combined effects of PFDM and rotation.  
\end{abstract}

\maketitle

\section{Introduction}
It is  widely believed that black holes typically emerge through the collapse of an immense star during the latter phase of its life-cycle \cite{os,penrose}. These stand as some of the most captivating revelations in the realm of science. Their presence has been definitively established through the identification of gravitational waves \cite{gw}, shadow images \cite{sh1,sh2,sh3,sh4,sh5}, tracking of stellar movements within the Galactic Center, and the precise measurement of the orbital trajectories of these stars. These methods offer a profound opportunity to explore the fundamental essence of black holes and the deep nature of spacetime  \cite{s1,s2,s3}
These intricate objects offer way for delving into various realms of physics, encompassing quantum gravity, thermodynamics, superconducting phase transitions, paramagnetism-ferromagnetism phase transitions, superfluidity, condensed matter physics, information theory, and the holographic hypothesis. One of the most captivating revelations within the domain of general relativity is the profound interplay between black holes and thermodynamics. In recent times, this relationship has undergone intensive scrutiny within the field of physics. The profound connection between black holes and thermodynamics can be traced back to the groundbreaking work of Hawking \cite{hawkingbh}. Nevertheless, astronomical observations compellingly suggest that black holes are enveloped by dark matter and that they possess a rotational aspect, necessitating the utilization of the Kerr spacetime geometry.

In a recent development, a novel approach for investigating the thermodynamic aspects of black holes has been presented in the work referenced as Ref. \cite{Wei:2022dzw}. In this paper, the authors have done a comprehensive examination of black holes by studying the topological classes of black hole thermodynamics employing the generalized off-shell free energy framework. As a consequence of this approach, the manifold of black hole solutions is categorized into three distinct topological classes, their differentiation arising from diverse topological numbers. Within this manuscript, our focus is directed towards an examination of the thermodynamic topological categories pertaining to both stationary and rotating black holes immersed in the framework of PFDM. To comprehend the impact of PFDM upon critical aspects such as black hole shadows, quasinormal modes, and deflection angles, a comprehensive reference can be found in the following Refs.\cite{pfdm0,pfdm1,pfdm2,pfdm3,pfdm4,pfdm5}. The topological approach to study the thermodynamics was used in recent works \cite{top1,top2,top3,top4,top5,top6} and, in particular, it was extended to rotating black holes in Ref. \cite{Wu:2022whe}. 
In what follows, we shall sketch a few concepts and provide an overview of the topological methodology introduced in \cite{Wei:2022dzw}. First of all, one has to use the notion of the generalized off-shell Helmholtz free energy, given by
\begin{equation}
\mathcal{F} = M -\frac{S}{\tau}.
\end{equation}
Once we have a black hole solution, several fundamental parameters come into play. These include the black hole mass denoted as $M$, the entropy represented by $S$, and an essential factor $\tau$, inversely related to the temperature of the encompassing cavity. Furthermore, as an integral component of this framework, a core vector $\phi$ can be introduced, as below \cite{Wei:2022dzw}
\begin{equation}
    \phi = \Big(\frac{\mathcal{\partial F}}{\partial r_{h}}\, , ~-C\cot\Theta\csc\Theta\Big) \, ,
\end{equation}
where $C$ is some arbitrary constant. In a broader context, this constant possesses the capacity to alter the orientation of the unit vector $n$. For ease of exposition, we will set $C = 1$. The two key parameters, $r_h$ and $\Theta$, conform to the conditions $0 < r_h < +\infty$ and $0\le \Theta\le \pi$, respectively. Notably, the component $\phi^\Theta$ exhibits divergence at $\Theta = 0,\pi$, thereby signifying an outward vector orientation in these regions. Utilizing Duan's theory of $\phi$-mapping topological currents, we can write the expression for a topological current as follows \cite{Duan:1979ucg,Duan} (see also \cite{Wei:2022dzw})
\begin{equation}
j^{\mu}=\frac{1}{2\pi}\epsilon^{\mu\nu\rho}\epsilon_{ab}\p_{\nu}n^{a}\p_{\rho}n^{b}\, , \qquad
\mu,\nu,\rho=0,1,2,
\end{equation}
where $\p_{\nu}= \p/\p{}x^{\nu}$ and $x^{\nu}=(\tau,~r_h,~\Theta)$. The unit vector $n$ reads as
$n = (n^r, n^\Theta)$, where $n^r = \phi^{r_h}/||\phi||$ and $n^\Theta = \phi^{\Theta}/||\phi||$.
It is simple to demonstrate that the topological current given above is conserved,
allowing one to easily deduce $\p_{\mu}j^{\mu} = 0$. It is then established that the topological
current $j^\mu$ is a $\delta$-function of the field configuration \cite{Duan:1979ucg,Duan}
\be
j^{\mu}=\delta^{2}(\phi)J^{\mu}\Big(\frac{\phi}{x}\Big)\, ,
\ee
where the 3-dimensional Jacobian $J^{\mu}\big(\phi/x\big)$ is defined as: $\epsilon^{ab}J^{\mu}
\big(\phi/x\big) = \epsilon^{\mu\nu\rho}\p_{\nu}\phi^a\p_{\rho}\phi^b$. It is simple to indicate
that $j^\mu$ equals to zero only when $\phi^a(x_i) = 0$, hence the topological number $W$ can
be determined as follows:
\be
W = \int_{\Sigma}j^{0}d^2x = \sum_{i=1}^{N}\beta_{i}\eta_{i} = \sum_{i=1}^{N}w_{i}\, ,
\ee
where $\beta_i$ is the positive Hopf index that counts the number of the loops of the vector
$\phi^a$ in the $\phi$-space when $x^{\mu}$ are around the zero point $z_i$, whilst $\eta_{i}=
\textit{sign}(J^{0}({\phi}/{x})_{z_i})=\pm 1$ is the Brouwer degree, and $w_{i}$ is the winding
number for the $i$-th zero point of $\phi$ that is contained in $\Sigma$. 

This paper is organized as follows: In the subsequent section, we study the topological classifications concerning the Kerr and Schwarzschild black holes within the context of PFDM. Moving forward, in sections III our focus shifts to an examination of the thermodynamic topological classes associated with Kerr and Kerr-AdS black holes in the presence of PFDM. We extend our analysis in section IV to encompass the Kerr-Newman black hole existing within the PFDM framework. Additionally, in section V, we elaborate on the impact induced by a nonlinear magnetic charge, specifically in the form of a rotating Hayward black hole within PFDM. Our paper culminates with a presentation of conclusions, outlined in section VI.

\section{Kerr Black Hole in PFDM Background}\label{sec2}
To begin, let us recall that the metric for a black hole in PFDM can be derived through the utilization of the action for the gravity theory minimally coupled to PFDM \cite{KerrDM}
\begin{eqnarray}
\mathcal{I}=\int dx^4\sqrt{-g}\left( \frac{1}{16\pi G} R+ \mathcal{L}_{DM}\right),
\end{eqnarray}
then from the Einstein's field equations obtained as 
\begin{eqnarray}
R_{\mu \nu}-\frac{1}{2}g_{\mu \nu}R =  8\pi G T_{\mu \nu}^{DM},
\end{eqnarray}
along with the effective energy density of PFDM that is given by
  \begin{equation}
  T_{t}^{\,t}=T_{r}^{\,r}=\frac{1}{8\pi}\frac{\alpha}{r^{3}}. 
  \end{equation}
One can then obtain the static solution and perform a \textcolor{black}{Newman–Janis} method to get the line element of the Kerr-like black hole in the dark matter background is
given as \cite{KerrDM}
\begin{eqnarray}\label{LEK}
ds^{2} &=&-\left(1-\frac{2mr-\alpha r \ln\left(\frac{r}{|\alpha|}\right)}{\Sigma}\right)dt^2+\frac{\Sigma}{\Delta_r}dr^2 \\
&+&\Sigma d\theta^2-2a\left(\frac{2mr-\alpha r \ln\left(\frac{r}{|\alpha|}\right)}{\Sigma}\right)dtd\phi\notag\\\notag
&&+\sin^2\theta\left(r^2+a^2+a^2\sin^2\theta\frac{2mr-\alpha r\ln\left(\frac{r}{|\alpha|}\right)}{\Sigma}\right)d\phi^2,
\end{eqnarray}
where
\begin{eqnarray}
\Delta_r &=& r^{2}-2mr+a^{2}+\alpha r\ln \left( \frac{r}{|\alpha |}\right),\\
\Sigma &=& r^{2}+a^{2}\cos ^{2}\theta. 
\end{eqnarray}
Here $m$ is the mass parameter, $a$  is  the rotation parameter, and $\alpha$ represents the dark matter parameter. For each chosen value $\alpha$ and the corresponding $a<k_c$\footnote{For reference the value of $k_c$ given by $k_c=\frac{\alpha}{2}\sqrt{\text{ProdcutLog}\left(2e^{-1+\frac{2}{\alpha}}\right)\left[2+\text{ProdcutLog}\left(2e^{-1+\frac{2}{\alpha}}\right)\right]}$.}, the line element \eqref{LEK} represents a black hole with two horizons, the inner horizon $r_-$ and the outer (Cauchy) horizon $r_h$, respectively \cite{rizwanatal}. For $a=k_c$, the line element represents an extremal black hole 
with one horizon\footnote{The size of extremal black hole horizon is given by
\begin{eqnarray}
    r_e=\frac{\alpha}{2} \text{ProductLog}\left(2e^{-1+\frac{2}{\alpha}}\right),\notag
\end{eqnarray}
where ProductLog$\left(x\right)$ represent Lambert $W$-function.}. The size of the black hole depends on the PFDM parameter $\alpha$. The mass $M$, entropy $S$, and the Hawking temperature $T$ of the Kerr black hole in PFDM background is given by \cite{MJP}
\begin{eqnarray}
M&=&m,\\
S&=&\pi\left(r_h^2+a^2\right),\\
T&=&\frac{r_h}{4\pi\left(r_h^2+a^2\right)}\left(1-\frac{a^2}{r^2_h}+\frac{\alpha}{r_h}\right).
\end{eqnarray}
The generalized off-shell free energy of the system in this case takes the form
\begin{eqnarray}\label{F}
\mathcal{F}&=&\frac{1}{2r_h}\left[r^2_h+a^2+\alpha r_h \ln\left(\frac{r_h}{|\alpha|}\right)\right]-\frac{\pi \left(r_h^2+a^2\right)}{\tau}.
\end{eqnarray}
Thus, the component of the vector field $\phi$ can be written as
\begin{eqnarray}\label{F}
\phi^{r_h}&=&\frac{1}{2r_h^2}\left(r^2_h+\alpha r_h-a^2\right)-\frac{2\pi r_h}{\tau},\\
\phi^\theta &=&-\cot{\Theta}\csc{\Theta}.
\end{eqnarray}
The inverse temperature parameter surrounding the Kerr black hole in PDFD can be written as 
\begin{eqnarray}
    \tau=\frac{4\pi r^3_h}{r^2_h+\alpha r_h-a^2}\equiv\mathcal{G}(r_h).
\end{eqnarray}
This shows the inverse temperature parameter $\tau$ (which is zero point of the $n$ vector field $\phi$) decreases with increasing $\alpha$. 
Now, to find the winding number $w_i$, and the topological number $W$, we use the approach given in \cite{wu2023topological} and verify our results by plotting the vector field $\phi$. Thus, we define a characterized complex function as
\begin{eqnarray}
    \mathcal{R}(z)=\frac{1}{\tau-\mathcal{G}(z)}.
\end{eqnarray}
The winding number $\omega_i$ is define as \cite{wu2023topological}
\begin{eqnarray} w_i=\frac{\text{Res}\mathcal{R}\left(z_i\right)}{|\text{Res}\mathcal{R}\left(z_i\right)|}\equiv \text{Sgn}\left[\text{Res}\mathcal{R}\left(z_i\right)\right],
\end{eqnarray}
where $\text{Res}\mathcal{R}\left(z_i\right)$ represents residue of  $\mathcal{R}\left(z_i\right)$ corresponding to the singular point $z_i$, Sgn$(x)$ represents sign function and $|z|$ represents the absolute value of $z$. 

\begin{figure*}[!ht]
 \includegraphics[width=7.5cm,height=7.5cm]{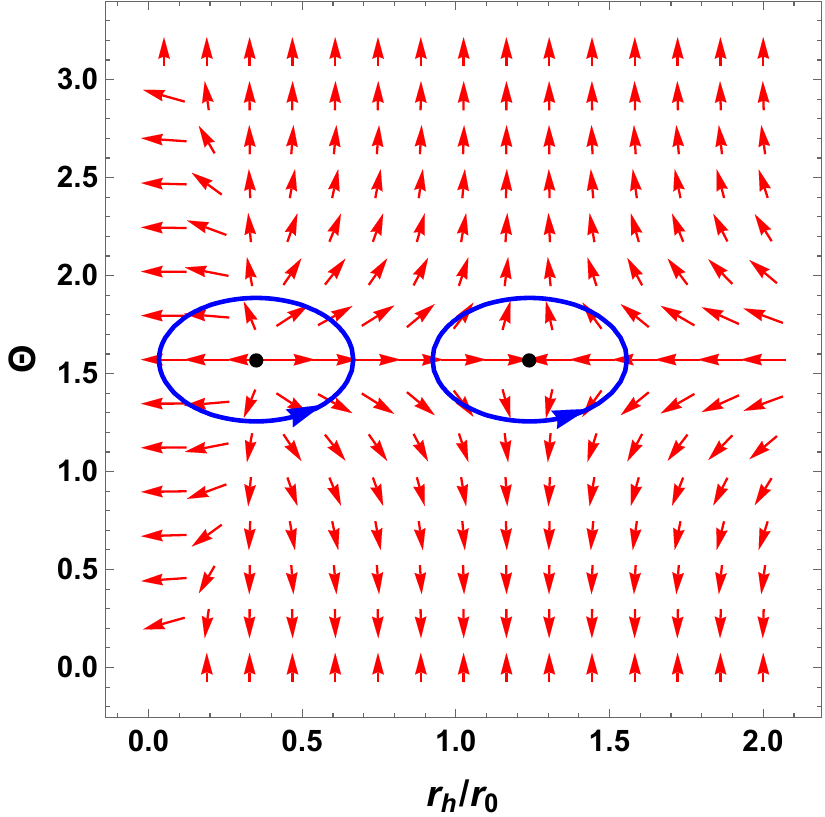}
 \includegraphics[width=7.5cm,height=7.5cm]{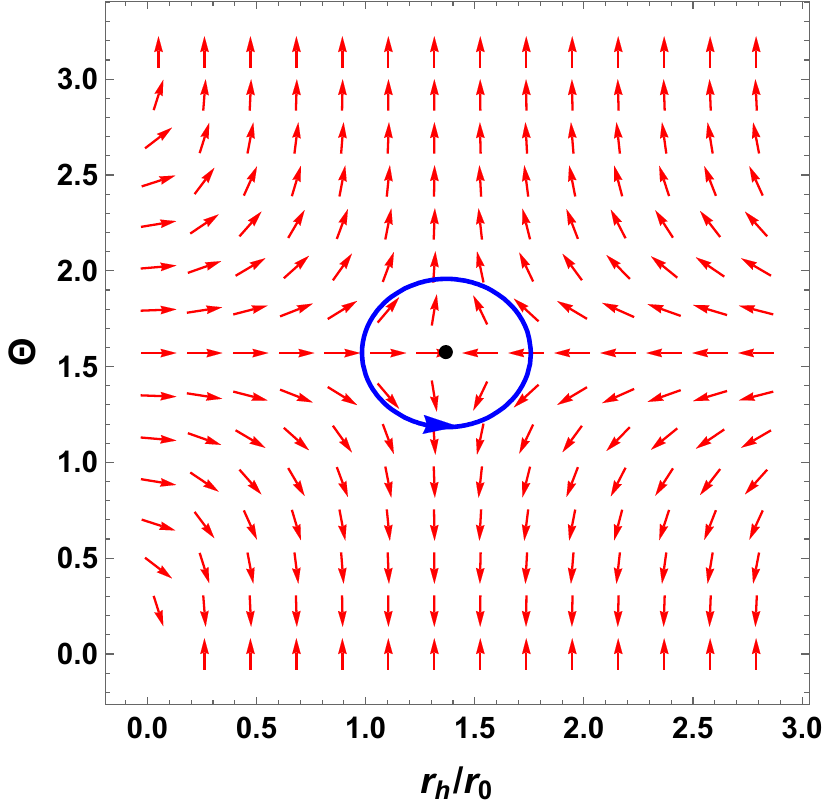}
  \caption{Left hand side: The red arrows represent unit vector field $n$ on a portion of the $r_h$-$\Theta$ plane plotted for the Kerr black hole in PFDM with $a/r_0=1/2$, $\tau/r_0=4\pi$ and $\alpha/r_0=1/2$. Right hand side: the plot of the vector field for the Schwarzschild black hole in PFDM with $\tau/r_0=4\pi$ and $\alpha/r_0=1/2$. The zero points are marked with black dots and are located at (left hand side) $(r_h/r_0,\Theta)=(0.35,\pi/2)$, and $(1.24,\pi/2)$; (right hand side)  $(r_h/r_0,\Theta)=(1.37,\pi/2)$.}
\label{nvfS&KBH}
  \end{figure*}

For the Kerr black hole in PFDM the characterized complex function takes the form
\begin{eqnarray}\label{CFKBH}
\mathcal{R}\left(z\right)&=&-\frac{z^2+\alpha z-a^2}{4\pi z^3-\tau\left(z^2+\alpha z-a^2\right)}\equiv-\frac{z^2+\alpha z-a^2}{\mathcal{A}\left(z\right)}.  
\end{eqnarray}
The roots analysis of $\mathcal{A}\left(z\right)=0$ shows\footnote{the cubic equation has discriminant $\delta=\tau^2\left[4a^2\left(\tau^2-108a^2\pi^2\right)+\alpha \tau\left(72a^2\pi+\alpha \tau+16 \pi \alpha^2\right)\right]$. If the discriminant is positive, then the equation has one negative and two positive real roots. If the discriminant is negative, then the equation has one negative  real and two complex roots. Further, if the discriminant is zero, then the equation has one negative and one positive root of multiplicity 2.}, for any choice of $\alpha$ with $a<k_c$ and $\tau<\tau_c$ where 
\begin{eqnarray}
    \tau_c=4\pi\frac{\left(-\alpha+\sqrt{3a^2+\alpha^2}\right)^3}{2a^2-\alpha\left(-\alpha+\sqrt{3a^2+\alpha^2}\right)},
\end{eqnarray}
there is one real negative and two complex roots. So, the off-shell condition cannot be satisfied. However, for $\tau_c<\tau$,
we have three singular points such that
\begin{eqnarray}
    z_1,z_2,z_3\in\mathbb{R},~ \text{and}~ 0<z_2<z_1~\text{whereas}~z_3<0.
\end{eqnarray}
So, the corresponding winding numbers and the topological number are 
\begin{eqnarray}
w_1=-1,~w_2=1~\text{and}~W=0.
\end{eqnarray}
\textcolor{black}{This shows that the winding numbers and the topological number for Kerr black hole in PFDM are the same as that of the Kerr black hole \cite{Wu:2022whe}.} This can also be seen from Fig. \ref{nvfS&KBH} (left hand side) which is plotted for unit vector field $n$ for the Kerr black hole in PFDM. The black dots represent the zero points of the vector field $\phi$. 

\begin{figure*}[!ht]
 \includegraphics[width=7.5cm,height=7.5cm]{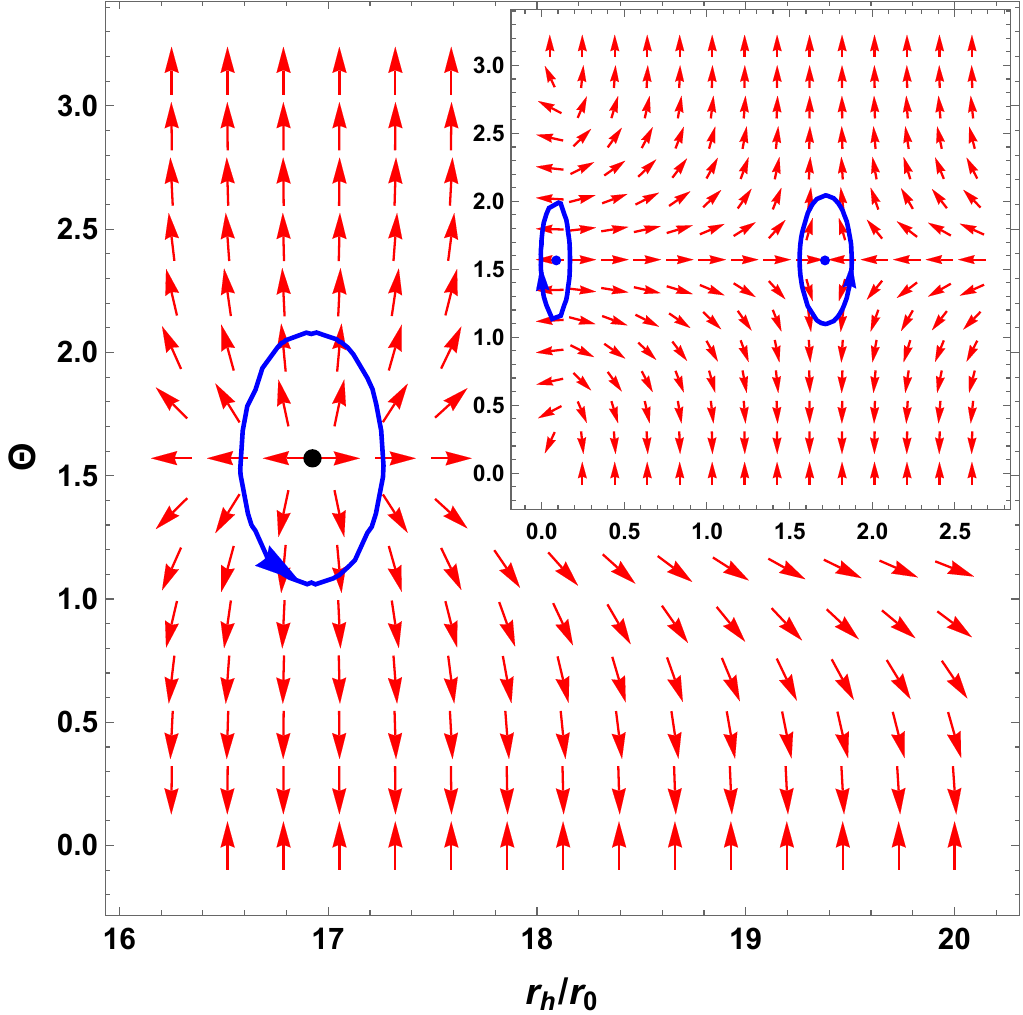}
 \includegraphics[width=7.5cm,height=7.5cm]{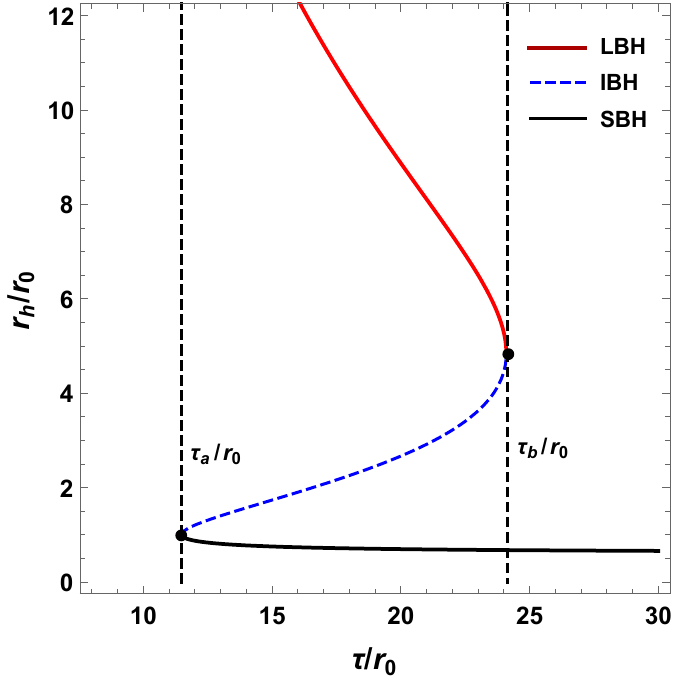}
  \caption{Left hand side: The $n$ vector field $\phi$ for  Kerr-AdS black hole in PFDM with $a/r_0=0.3$, $\tau/r_0=4\pi$, $\alpha/r_0=1/2$ and $Pr^2_0=0.0022$ is plotted. There are three zero points (two are shown in the inset which is plotted for small values of $r_h/r_0$) and are located at $(r_h/r_0,\Theta)=(0.08,\pi/2)$, $(1.72,\pi/2)$, and  $(r_h/r_0,\Theta)=(16.92,\pi/2)$. Right hand side: The zero points of $\phi^{r_h}$ with $a=r_0$, $\tau/r_0=4\pi$, $\alpha/r_0=1$ and $Pr^2_0=0.0022$ are for Kerr-Ads Black hole in PFDM is shown. The red solid, blue dashed, and black solid lines are for the large, intermediate, and small black hole, respectively. Further, for Kerr-Ads black hole in PFDM there is a generation point as well as an annihilation point represented by black dots.}
\label{KAdsDMSILBH}
  \end{figure*}

\begin{figure}
\centering
\includegraphics[width=7.5cm,height=7.5cm]{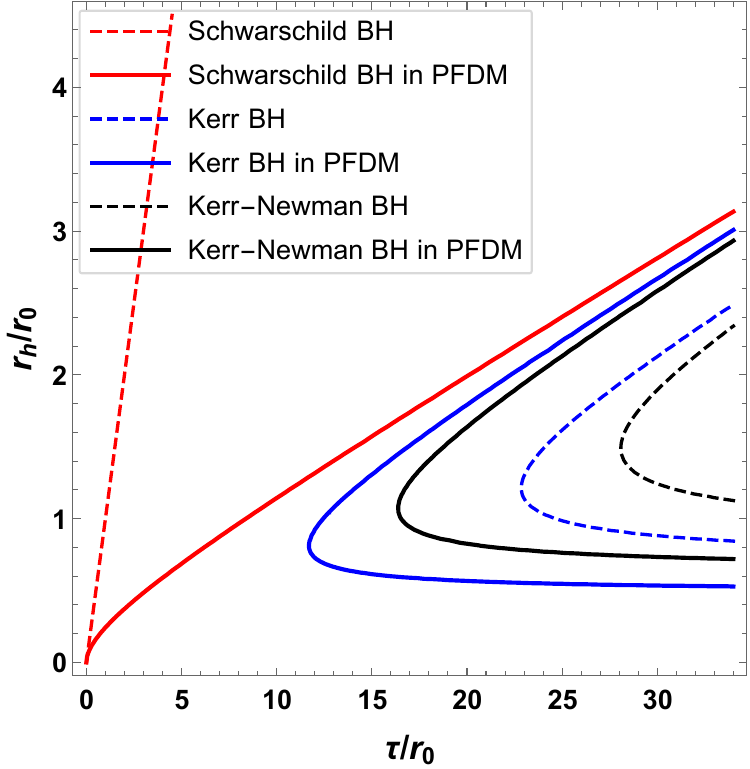}
\caption{The zero points of vector field $\phi$  are plotted in $r_h$-$\tau$ plane for a family of black holes spacetimes. The figure shows that the rotation and electric charge affects the topological behavior of spacetime as the charged and rotating black holes have generation points and different topological numbers from Schwarzschild black holes.}
\label{zeroallblackhole}
\vspace{1.7cm}
\end{figure}

\begin{figure}[h!]
 \includegraphics[width=7.5 cm,height=7.5 cm]{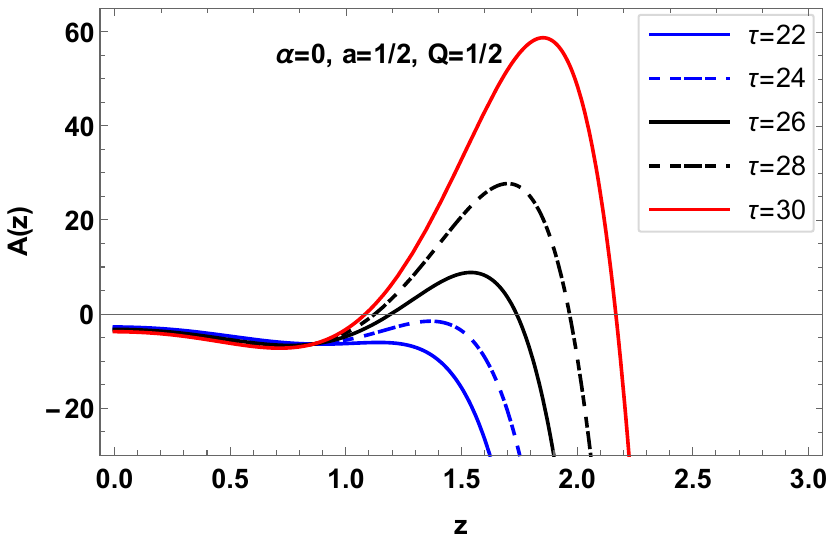}
  \caption{The $\mathcal{A}(z)$ [defined by Eq. (\ref{azRHBH})] is plotted for different mentioned values for rotating Hayward BH in the absence of PFDM. We see from this case that for $\tau_c<\tau$, $\mathcal{A}(z)=0$ has two real positive roots.}
  \label{AzfHBH}
  \vspace{0.7cm}
  \end{figure}

  \begin{figure}[h!]
 \includegraphics[width=7.5cm,height=7.5 cm]{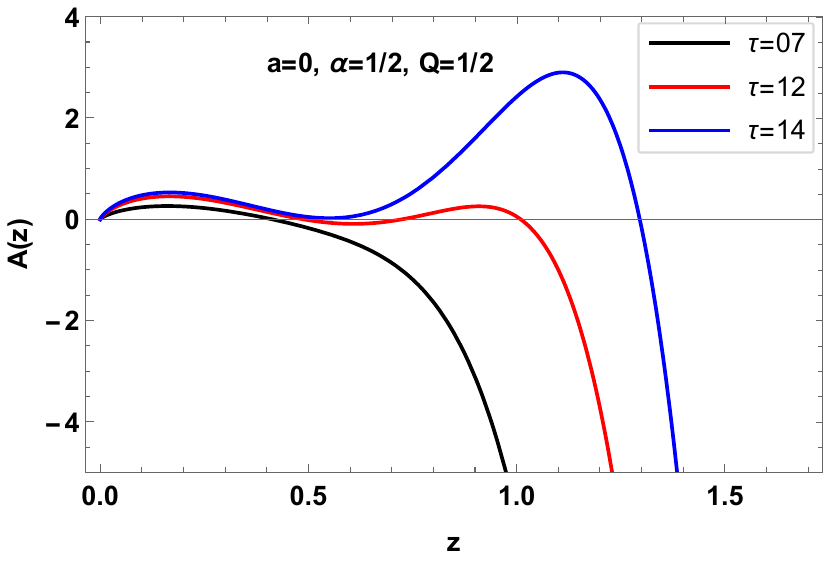}
  \caption{The plot of the $\mathcal{A}(z)$ [define by Eq. (\ref{azfnRHBH})] for the static Hayward BH in the presence of PFDM. This shows that just like Kerr-Ads black hole, for the non-rotating Hayward BH in PFDM background, there are two critical values of $\tau$, $\tau_a$ and $\tau_b$ such that if $\tau\in(\tau_a,\tau_b)$, the equation $\mathcal{A}(z)=0$ has three real positive roots and one real root otherwise. Similar behavior is seen for other choices of the parameter $\alpha$, $a$, and $Q$. }
  \label{AzfHBHDMB}
  \end{figure}

  \begin{figure}[h!]
 \includegraphics[width=7.5cm,height=7.5 cm]{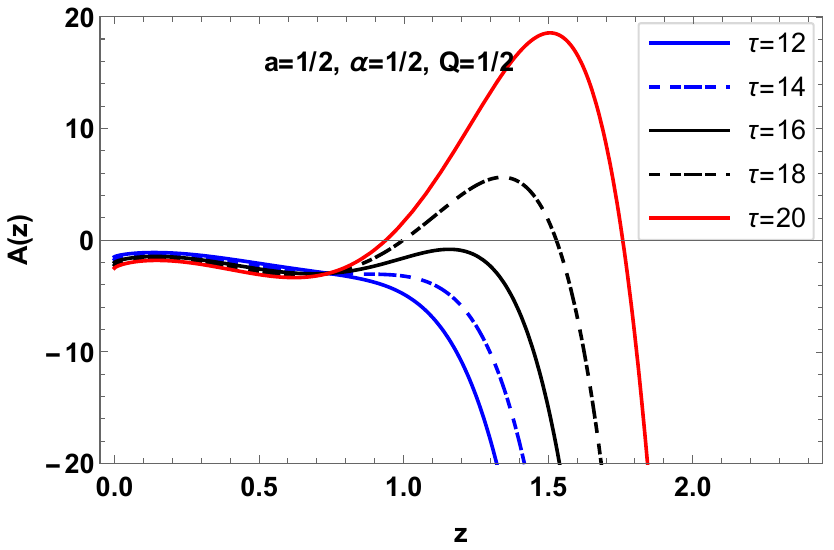}
  \caption{The plot of $\mathcal{A}(z)$ [defined by Eq. (\ref{azfGHBH})] for the general case of rotating Hayward BH in the presence of PFDM. Similar result we find here for $\tau_c<\tau$, $\mathcal{A}(z)=0$ has two real positive roots and in both cases, the rotating Hayward and rotating Hayward BH in PFDM background have same winding numbers $w_1=-1$ and $w_2=1$ with topological number $W=0$.  }
  \label{AzfRHBHDMB}
  \end{figure}

\subsection{Schwarzschild Black Hole in Dark Matter Background}
As a special case of the Kerr black hole in PFDM we can obtain the static line element; namely, the Schwarzschild black hole in PFDM with the line element
\begin{eqnarray}\label{SBHPFDM}
ds^{2} &=&-f(r)dt^2+\frac{1}{f(r)}dr^2+r^2(d\theta^2+\sin^2\theta d\phi^2),
\end{eqnarray}%
with
\begin{eqnarray}\label{fofBH}
f(r)=1-\frac{2m}{r}+\frac{\alpha}{r} \ln\left( \frac{r}{|\alpha |}\right),
\end{eqnarray}%
where $m$ representing the mass and $\alpha$ as the PFDM parameter. Note, that for every choice of the parameter $\alpha$, the line element \eqref{SBHPFDM} represents a black hole with one horizon called event horizon represented by $r_h$. Further, the size of the event horizon increases with increasing the dark matter parameter $\alpha$ \cite{rizwanatal}. The thermodynamic quantities for the Schwarzschild black hole in PFDM background are given as \cite{MJP}
\begin{equation}
    M= m,\quad S=\pi r_h^2, \quad  T=\frac{1}{4\pi r_h}\left(1+\frac{\alpha}{r_h}\right).
\end{equation}

In the absence of the PFDM, that is, for $\alpha=0$, the line element and the thermodynamical quantities reduce to that of the Schwarzschild black hole. For the Schwarzschild black hole in PFDM, the generalized off-shell free energy takes the form
\begin{eqnarray}\label{F}
\mathcal{F}&=&\frac{1}{2}\left[r_h+\alpha \ln\left( \frac{r_h}{|\alpha |}\right)\right]-\frac{\pi r_h^2}{\tau}.
\end{eqnarray}
Further, the components of the vector field $\phi$ are
\begin{eqnarray}
\phi^{r_h}&=&\frac{1}{2}\left(1+\frac{\alpha }{r_h}\right)-\frac{2\pi r_h}{\tau},\\
\quad \phi^\theta &=&-\cot{\Theta}\csc{\Theta},
\end{eqnarray}
which gives the values of the inverse temperature parameter ($\tau$) of the cavity surrounding the black hole as
\begin{eqnarray}
    \tau=\frac{4\pi r_h^2}{r_h+\alpha}.
\end{eqnarray}

This shows that $\tau$ decreases  with increasing the dark matter parameter $\alpha$. Further, in the absence of PFDM $(\alpha=0)$, the result reduces to that of the Schwarzschild black hole \cite{Wei:2022dzw}. Now, the characterized complex function is given as
\begin{eqnarray}\label{Rsz}
\mathcal{R}\left(z\right)&=&-\frac{z+\alpha}{4\pi z^2-\tau\left(z+\alpha\right)}. 
\end{eqnarray} 

The function $\mathcal{R}\left(z\right)$ for all values of $\alpha$ has two singular points say, $z_1$ and $z_2$, such that
\begin{eqnarray}
    z_1, z_2\in\mathbb{R},~ \text{and} ~ z_2<0 ~  \text{whereas}  ~0<z_1.
\end{eqnarray}
For the negative real root $z_2$, the off-shell condition cannot be satisfied and therefore, we only consider the positive singular value $z_1$ which gives the corresponding winding number $w_1$ and the topological number as
\begin{eqnarray}
    w_1=-1~ \text{and}~W=-1.
\end{eqnarray}
This indicates that the winding number and topological classes for the Schwarzschild black hole within the PFDM framework align with those of the Schwarzschild black hole  \cite{Wei:2022dzw}. Conversely, they diverge from the corresponding values for the Kerr black hole in the PFDM context. 
Hence, the rotation of spacetime within the context of PFDM exerts an influence on its topological characteristics. This assertion finds support in the graphical representation of the $n$ vector field, as depicted in Fig.~\ref{nvfS&KBH} (right hand side).

\section{Kerr-AdS Black Hole in Dark Matter Background}
In this section, we discuss the effect of PFDM on topological classes of Kerr-Ads spacetime. The line element of the Kerr-Ads black hole in the dark matter background is given as \cite{KerrDM}
\begin{eqnarray}\label{LE}\notag
ds^{2} &=&-\frac{\Delta_r}{\Xi\Sigma}\left(dt-a\sin^2\theta d\phi\right)^2+\frac{\Sigma }{\Delta }dr^{2}+\frac{\Sigma}{\Delta_\theta} d\theta ^{2}\\
&+&\frac{\Delta_\theta \sin^2\theta}{\Xi\Sigma} \left[adt-\left(r^2+a^2\right)d\phi\right]^2 ,
\end{eqnarray}%
where%
\begin{eqnarray}\notag
\Delta_r &=& r^{2}-2mr+a^{2}-\frac{\Lambda}{3}r^2\left(r^2+a^2\right)+\alpha r\ln \left( \frac{r}{|\alpha |}\right), \\\notag
\Delta_\theta &=&1+\frac{\Lambda}{3}a^2\cos^2\theta, \\
\Sigma&=& r^{2}+a^{2}\cos ^{2}\theta \quad \Xi=1+\frac{\Lambda}{3}a^2.
\end{eqnarray}%
Here $m$ and $a$ are the mass and angular momentum per unit mass parameters of the black hole and $\alpha$ and $\Lambda$ are the PFDM parameter and cosmological constant, respectively. 

In the absence of PFDM ($\alpha =0$), the line element %
\eqref{LE} represents a Kerr-AdS black hole. The location of the black hole horizons can be obtained by solving the horizon equation 
\begin{equation}  \label{Horizneq}
\Delta_r =r^{2}-2mr+a^{2}-\frac{\Lambda}{3}r^2\left(r^2+a^2\right)+\alpha r\ln \left( \frac{r}{|\alpha |}\right) =0.
\end{equation}%
The mass and other thermodynamical quantities are given by given by \cite{MJP}
\begin{eqnarray}
M&=&\frac{m}{\Xi^2},\\
S&=&\frac{\pi\left(r_h^2+a^2\right)}{\Xi},\\
T&=&\frac{r_h}{4\pi\Xi\left(r_h^2+a^2\right)}\left[1-\frac{a^2}{r^2_h}-\frac{\Lambda}{3}\left(3r^2_h+a^2\right)+\frac{\alpha}{r_h}\right].
\end{eqnarray}
Using these thermodynamical quantities generalized off-shell free energy of the system can be obtained as
\begin{equation}
\mathcal{F}=\frac{1}{2\tau r_h\left(3-8\pi a^2P\right)^2}
\Big[3 \left(r_h^2+a^2\right)\mathcal{G}+9 \alpha r_h \tau \ln \left(\frac{r_h}{\left| \alpha \right| }\right)\Big],
\end{equation}
where 
\begin{eqnarray}
    \mathcal{G}=2 \pi r_h \left(8 \pi P a^2 +4\tau P r_h-3\right)+ 3 \tau.
\end{eqnarray}
which gives the the components of $n$ vector field as
\begin{eqnarray}
   \phi^{r_h}&=&\frac{6\pi r^2_h\left\{8\pi a^2r_h P-3r_h+2\tau P\left(a^2+3r^2_h\right)\right\}}{2r^2_h\tau\left(3-8\pi a^2 P\right)^2}\nonumber\\
   &&+\frac{9\tau\left(r^2_h+\alpha r_h-a^2\right)}{2\tau r^2_h\left(3-8\pi a^2P\right)},\\
   \phi^\theta&=&-\cot\Theta \csc\Theta.
\end{eqnarray}
Thus, the inverse temperature for the Kerr-Ads black hole in PFDM as 
\begin{eqnarray}
    \tau=\frac{4 \pi r_h^3 \left(3 -8 \pi P a^2 \right)}{3\left(r_h^2+ \alpha  r_h-a^2\right)+8 \pi P a^2 r_h^2\left(a^2+ 3r^2_h\right)}.
\end{eqnarray}
The characterized function in this case takes the form
\begin{eqnarray}
    \mathcal{R}(z)\equiv-\frac{3\left(z^2+\alpha z-a^2\right)+8\pi Pa^2z\left(a^2+3z^2\right)}{\mathcal{A}(z)},
\end{eqnarray}
where
\begin{eqnarray}
    \mathcal{A}(z)&=&4\pi z^3\left(3-8\pi P a^2\right)-\tau\Big\{3\left(z^2+\alpha z-a^2\right)\nonumber\\
    &+& 8\pi P a^2 z \left(a^2+3z^2\right)\Big\}.
\end{eqnarray}
The root analysis of $\mathcal{A}(z)=0$, shows that the for any chosen $a$ and $\tau<\tau_a$ or $\tau_b<\tau$ there is only one real positive root and hence the corresponding winding and topological number is $1$. Further, for $\tau \in (\tau_a,\tau_b)$ there are three positive and one negative root, and the corresponding winding numbers are $w_1=1,~w_2=-1,~w_3=1$ and hence again we have same topological number $W=1$. This result can be seen from the Fig. \ref{KAdsDMSILBH} (left hand side) in which vector field $\phi$ is plotted for $\tau\in(\tau_a,\tau_b)$ which has three zero points with the winding number $w_1=1,~w_2=-1,~w_3=1$ 
with the topological number $W=1$. The zero point of the vector field is plotted in  Fig. \ref{KAdsDMSILBH} (right hand side) which divides the black hole into a small, intermediate, and large black hole with one generation and one annihilation point represented by black dots. So, the Kerr-AdS black hole in PFDM has different topological behavior (with topological $W=1$) than that of the Kerr black hole in PFDM (with topological $W=0$).

\begin{figure*}[h!]
 \includegraphics[width=7.5cm,height=7.5 cm]{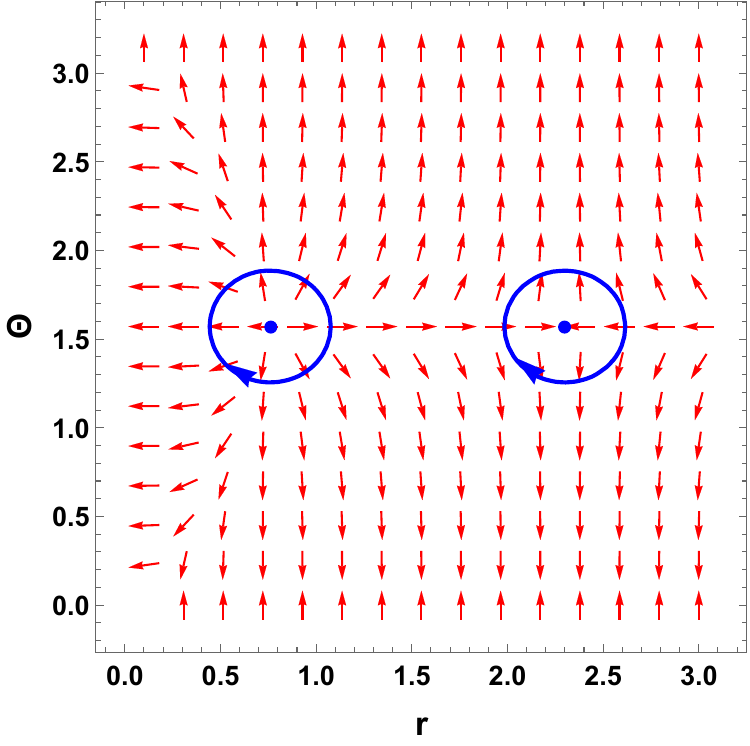}
 \includegraphics[width=7.5cm,height=7.5 cm]{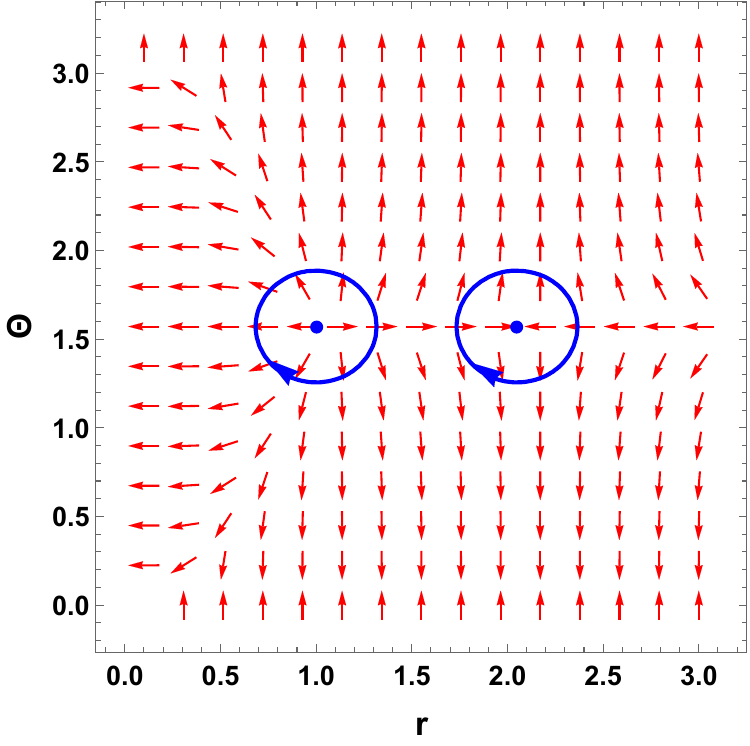}
 \includegraphics[width=7.5cm,height=7.5 cm]{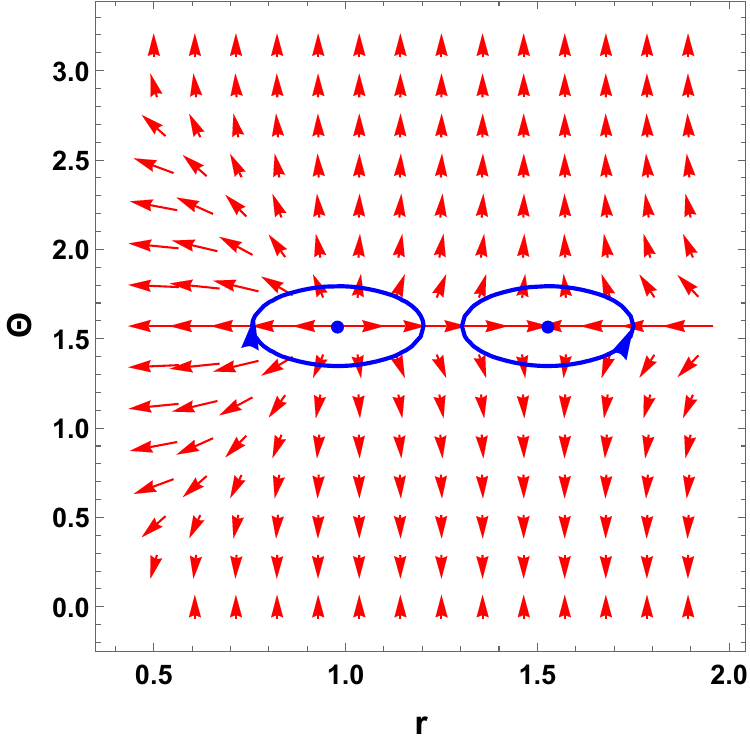}
 \includegraphics[width=7.5cm,height=7.5 cm]{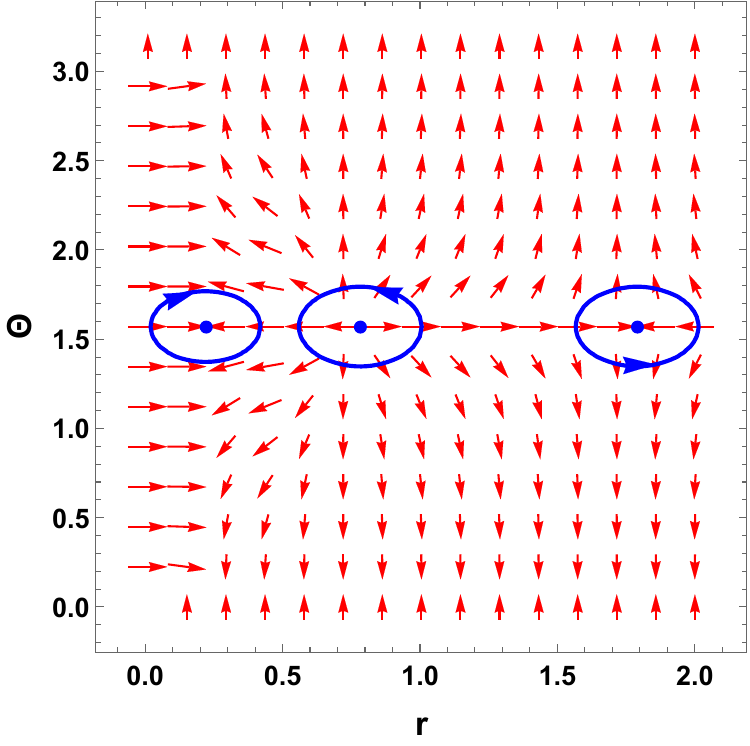}
  \caption{The unit vector field $\phi$ is plotted for the class of black holes with the magnetic charge. Top Left  Panel: The field for Hayward black hole (with $a=0,~Q=1/2,~\tau=30$) which shows that Hayward black hole has $2$ zero points. Top Right Panel: The field  for rotating Hayward black hole (with  $a=1/2,~Q=1/2,~\tau=30$) which has $2$ zero points.  Left bottom Panel: The field has plotted for the rotating Hayward black hole in PFDM (with $a=1/2,~Q=1/2,~\alpha=1/2,~\tau=18$). Right bottom: The field is plotted for Hayward black hole in PFDM (with $a=0,~Q=1/2,~\alpha=1/3,~\tau=20$) which shows unlike other black holes the Hayward black hole in PFDM has three zero points. A similar behavior is seen for other choices of parameters.}
  \label{vfHBHs}
  \end{figure*}

\begin{figure}[h!]
 \includegraphics[width=7.5 cm,height=7.5 cm]{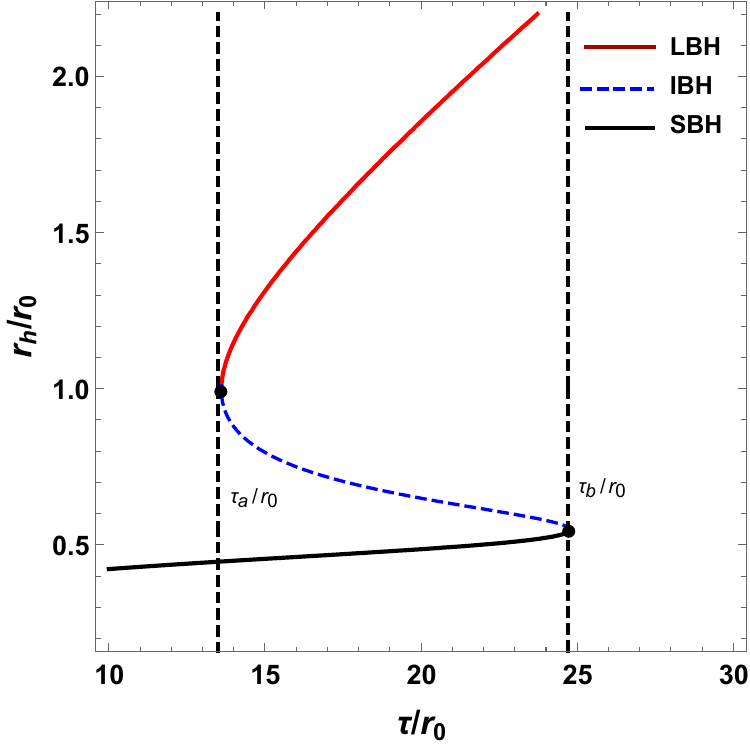}
  \caption{The zero point of $\phi^{r_h}$ for Hayward black hole in PFDM is plotted. The red solid, blue dashed, and black solid lines representing the large, intermediate, and small black hole, respectively. }\label{zeropHBHDM}
  \end{figure}

 \begin{figure}[h!]
 \includegraphics[width=7.5 cm,height=7.5 cm]{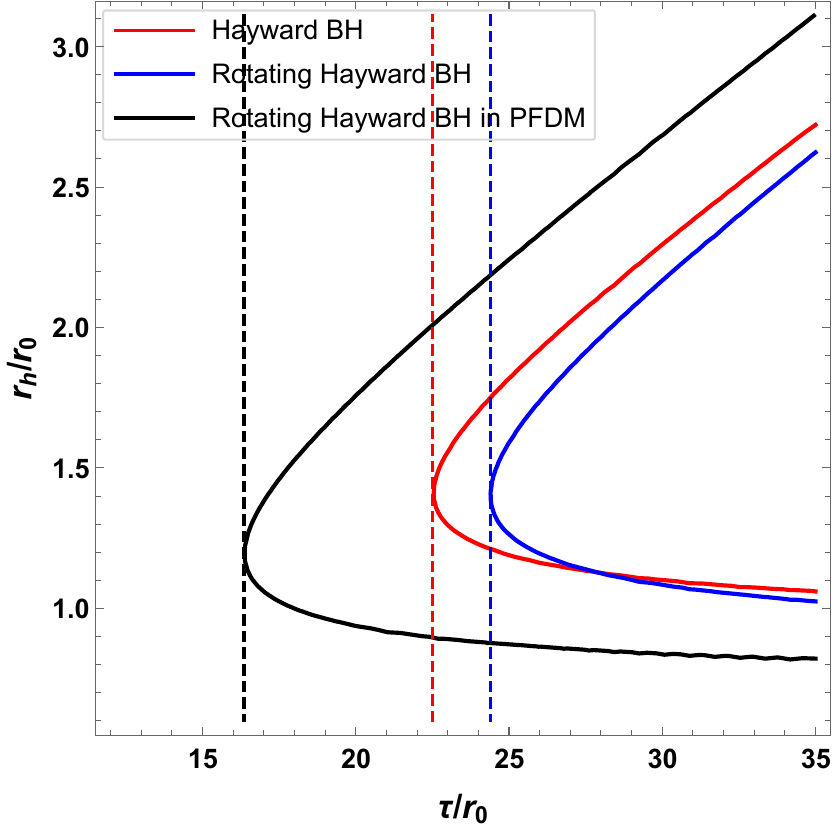}
  \caption{
  The zero point of $\phi^{r_h}$ for class of black hole is plotted which shows that the black holes has one generation point and no annihilation point.}
  \label{zpallBHs}
  \end{figure} 

\section{The role of electric charge: Kerr-Newman black hole in dark matter background}
Let's now direct our focus towards understanding the importance of the electric charge concerning the topological numbers of thermodynamics. To find the spacetime geometry, one can start from the action for the gravity theory minimally coupled with gauge field in PFDM reads as \cite{Das:2020yxw}
\begin{eqnarray}
\mathcal{I}=\int dx^4\sqrt{-g}\left( \frac{1}{16\pi G} R+\frac{1}{4} F^{\mu \nu}F_{\mu \nu} + \mathcal{L}_{DM}\right).
\end{eqnarray}
In the last equation $g$ = det($g_{ab}$) is the determinant of the metric tensor, $R$ is the Ricci scalar, further $G$ is Newton's gravitational constant, $F_{\mu \nu}= \nabla_\mu A_\nu -\nabla_\nu A_\mu$ is electromagnetic field tensor and $\mathcal{L}_{DM}$ is the Lagrangian density for PFDM. Upon using the variation for the action principle, one can obtain the Einstein field equations \cite{Das:2020yxw}
\begin{eqnarray}
R_{\mu \nu}-\frac{1}{2}g_{\mu \nu}R = 8\pi G (T_{\mu \nu}^M +T_{\mu \nu}^{DM}), 
\end{eqnarray}
along with
\begin{eqnarray}
F^{\mu \nu}_{;\nu}=0, \nonumber \\
F^{\mu \nu ;\alpha} +F^{\nu \alpha ;\mu}+F^{\alpha \mu ; \nu} =0.
\end{eqnarray} 
We have $T_{\mu \nu}^M$, which is the energy-momentum tensor for ordinary matter and $T_{\mu \nu}^{DM}$ that describes the energy-momentum tensor for PFDM 
 \begin{eqnarray}
T^\mu_\mu =g^{\mu \nu} T_{\mu \nu}, \nonumber \\
T^t_t = -\rho, \;\;\; T^r_r = T^{\theta}_{\theta} = T^{\phi}_{\phi} = P.
\end{eqnarray} 

It was shown that the line element of the Kerr-Newman like black hole  in the dark matter background is
given as \cite{Das:2020yxw}
\begin{eqnarray}\label{LEKNPFDM}\notag
ds^{2} &=&-\frac{\Delta}{\Sigma}\left(dt-a\sin^2\theta d\phi\right)^2+\frac{\Sigma }{\Delta }dr^{2}+\Sigma d\theta ^{2}\\
&+&\frac{ \sin^2\theta}{\Sigma} \left[adt-\left(r^2+a^2\right)d\phi\right]^2 ,
\end{eqnarray}%
where%
\begin{eqnarray}\notag
\Delta &=& r^{2}-2mr+Q^2+a^{2}+\alpha r\ln \left( \frac{r}{|\alpha |}\right), \\
\Sigma&=& r^{2}+a^{2}\cos ^{2}\theta. 
\end{eqnarray}%
Here $Q$ is the electric charge of the black hole. The root analysis of the horizon equation $\Delta=0$ shows that for any choice of $\alpha$  if the rotation parameter $a$ and charge parameter $Q$ are in the limit $a^2+Q^2<k^2_c$, the line element \eqref{LEKNPFDM} represents a black hole with three horizons and a naked singularity otherwise. The thermodynamic quantities for this spacetime can be written as
\begin{eqnarray}
    M&=&m,\\
    S&=&{\pi \left(r^2+a^2\right)},\\
    T&=& \frac{\Delta'(r_h)}{4 \pi (r_h^2+a^2)}
\end{eqnarray}
For this black hole, the off-shell free energy can be obtained as
\begin{eqnarray}
    \mathcal{F}&=&\frac{a^2+\alpha r_h \ln \left(\frac{r_h}{\left| \alpha \right| }\right)+Q^2+r^2_h}{2r_h}-\frac{\pi  \left(a^2+r^2_h\right)}{\tau}
\end{eqnarray}
This gives the components of the $n$ vector field as
\begin{eqnarray}
    \phi^{r_h}&=&\frac{1}{2r^2_h}\left(r^2_h+\alpha r_h-a^2-Q^2- \frac{4\pi r^3_h}{\tau}\right),\\
\phi^\theta &=&-\cot\Theta \csc\Theta,
\end{eqnarray}
and
\begin{eqnarray}
    \tau=\frac{4\pi r^3_h}{r^2_h+\alpha r_h-a^2-Q^2}
\end{eqnarray}
To find the winding number we define 
\begin{eqnarray}
    \mathcal{R}(z)=-\frac{z^2+\alpha z-a^2-Q^2}{4\pi z^3-\tau \left(z^2+\alpha z-a^2-Q^2\right)},
\end{eqnarray}
which has the same form as that of the Kerr black hole in PFDM [given by Eq. \eqref{CFKBH}] instead of $a^2\to a^2+Q^2$. So, a similar analysis as for the Kerr black hole in section \ref{sec2}, shows that for any choice of $\alpha$ with $a^2+Q^2<k^2_c$ and $\tau_c<\tau$, where
\begin{eqnarray}
    \tau_c=4\pi \frac{\Big(-\alpha+\sqrt{3(a^2+Q^2)+\alpha^2}\Big)^3}{2a^2-\alpha \Big(-\alpha+\sqrt{3(a^2+Q^2)+\alpha^2}\Big)},
\end{eqnarray}
the winding number and the topological number for the Kerr-Newman black hole in PFDM are 
\begin{eqnarray}
    w_1=-1,~w_2=1~\text{and}~W=0.
\end{eqnarray}

\subsection{The Reissner-Nordstr$\Ddot{o}$m black hole in
PFDM}
It has the line element 
\begin{eqnarray}\label{RNBHPFDM}
ds^{2} &=&-f(r)dt^2+\frac{1}{f(r)}dr^2+r^2(d\theta^2+\sin^2\theta d\phi^2),
\end{eqnarray}%
with
\begin{eqnarray}\label{fofBH}
f(r)=1-\frac{2m}{r}+\frac{Q^2}{r^2}+\frac{\alpha}{r} \ln\left( \frac{r}{|\alpha |}\right),
\end{eqnarray}%
The results for the Reissner-Nordstr$\Ddot{o}$m black hole in PFDM can be obtained as from the Kerr-Newman Black hole, which shows that for $Q<k_c$, and $\tau<\tau_c$, the winding number, and the topological number are
\begin{eqnarray}
    w_1=-1,~w_2=1~\text{and}~W=0.
\end{eqnarray}
The zero point $\tau$ of the vector field $\phi^{r_h}$ is plotted in Fig. \ref{zeroallblackhole} for the family of black holes which shows the different topological classes with winding and the topological number. For instant, the Schwarschild black hole and the Schwarschild black hole in PFDM have a topological number $-1$ with no generation/annihilation point whereas other black holes have a topological number equal to $0$ with one generation point. 

\section{The role of nonlinear magnetic charge: Rotating Hayward black hole in dark matter background}

In this section, we shall investigate the role played by the magnetic charge. In order to find the spacetime geometry, one has to start from the action for the gravity theory minimally coupled with gauge field in PFDM reads with the Einstein gravity coupled to a nonlinear electromagnetic field in the presence of the perfect fluid dark matter. One can show that the following equations can be obtained \cite{Ma:2020dhv}
  \begin{gather}
  G_{\mu}^{\,\nu}=2\left(\frac{\partial\mathcal{L}\left(F\right)}{\partial F}F_{\mu\lambda}F^{\nu\lambda}-\delta_{\mu}^{\,\nu}\mathcal{L}\right)+8\pi T_{\mu}^{\,\nu}\left(\mathrm{DM}\right),\label{eq:EM1}\\
  \nabla_{\mu}\left(\frac{\partial\mathcal{L}\left(F\right)}{\partial F}F^{\nu\mu}\right)=0.\label{eq:EM2}
  \end{gather}
  In which $F_{\mu\nu}=2\nabla_{[\mu}A_{\nu]}$ where $\mathcal{L}$ is a function of $F\equiv\frac{1}{4}F_{\mu\nu}F^{\mu\nu}$ and it is given by \cite{Ma:2020dhv}
  \begin{equation}
  \mathcal{L}\left(F\right)=\frac{3M}{\vert Q_m\vert^{3}}\frac{\left(2Q_m^{2}F \right)^{\frac{3}{2}} }{\left(1+ \left(2Q_m^{2}F \right)^{\frac{3}{4}}   \right)^{2} }.
  \end{equation}
  
It is straightforward to show that the line element of the rotating Hayward like black hole  in the dark matter background is
given as \cite{Ma:2020dhv}
\begin{eqnarray}\notag
ds^{2} &=&-\frac{\Delta}{\Sigma}\left(dt-a\sin^2\theta d\phi\right)^2+\frac{\Sigma }{\Delta }dr^{2}+\Sigma d\theta ^{2}\\
&+&\frac{ \sin^2\theta}{\Sigma} \left[adt-\left(r^2+a^2\right)d\phi\right]^2 ,
\end{eqnarray}%
where%
\begin{eqnarray}\notag
\Delta &=& r^{2}-\frac{2 m r^4}{r^3+Q_m^3}+a^{2}+\alpha r\ln \left( \frac{r}{|\alpha |}\right), \\
\Sigma&=& r^{2}+a^{2}\cos ^{2}\theta 
\end{eqnarray}%
Here $Q$ is the electric charge of the black hole. The mass $M$, magnetic charge $Q_m$, entropy $S$, and the Hawking temperature $T$ of the black hole are
\begin{eqnarray}
M&=&m,\\
S&=&=\pi \left[r_h^2+a^2-\frac{2 Q_m^3}{r_h}\left(1+\frac{a^2}{3 r_h^2}\right) \right], \\
T&=& \frac{\Delta'(r)}{4 \pi (r_h^2+a^2)}|_{r_h}.
\end{eqnarray}
Using the same method as that used for the previous section we can find the components of the $n$ vector field as
\begin{eqnarray}
    \phi^{r_h}&=&\frac{1}{2\tau r^5_h}\Big[r^3_h\left\{\tau\left(r^2_h+\alpha r_h-a^2\right)-4\pi r^3_h\right\}\nonumber\\
    &-&Q_m^3\tau \Big(2r^2_h+4a^2+\alpha r_h\left(1-3\ln\left({\frac{r_h}{|\alpha|}}\right)\right)\nonumber\\ 
    &+&\frac{4\pi r_h}{\tau}\left(r^2+a^2\right)\Big)\Big],\\
    \phi^\theta&=&-\cot\Theta\csc\Theta,
\end{eqnarray}
and the inverse temperature parameter is given by
\begin{eqnarray}
\tau=4\pi\frac{r^6_h+Q_m^3r_h\left(r^2_h+a^2\right)}{r^3_h\left(r^2_h+\alpha r_h-a^2\right)-Q_m^3\mathcal{H}\left(r_h\right)},
\end{eqnarray}
where
\begin{eqnarray}
    \mathcal{H}(r_h)=4a^2+2r^2_h-\alpha r_h+3\alpha r_h\ln\left(\frac{r_h}{|\alpha|}\right).
\end{eqnarray}
The complex function $\mathcal{R}(z)$ takes the form
\begin{eqnarray}
    \mathcal{R}(z)\equiv-\frac{z^3\left(z^2+\alpha z-a^2\right)-Q_m^3\mathcal{H}(z)}{\mathcal{A}(z)}
\end{eqnarray}
where
\begin{eqnarray}\label{azfGHBH}\notag
\mathcal{A}(z)&=&4\pi\left\{z^6+Q^3_mz\left(z^2+a^2\right)\right\}\\
&-&\tau\left\{z^3\left(z^2+\alpha z-a^2\right)-Q_m^3\mathcal{H}(z)\right\}. 
\end{eqnarray}
For the analysis of complex function $\mathcal{R}(z)$ and hence the winding number, we have plotted $\mathcal{A}(z)$ for rotating Hayward black hole in PFDM [define by Eq.\eqref{azfGHBH}] in Fig. \ref{AzfRHBHDMB}. This figure shows that for $\tau_c<\tau$, the equation $\mathcal{A}(z)$ has two positive real roots say $z_1$ and $z_2$ with $z_1<z_2$ such that the corresponding winding number are $w_1=-1$ and $w_2=1$ with topolgoical number $W=0$\footnote{the similar behaviour is seen for other choices of the parameters $a,~Q$ and $\alpha$.}. The result can be verified from the vector field plot in Fig. \ref{vfHBHs} (bottom left panel). Further, from the graph of zero point  $\phi^{r_h}$ in Fig. \ref{zpallBHs} we can see that the rotating black hole has $1$ or $0$ point but no annihilation point.  

\subsection{Static solution: Hayward black hole in the absence of PFDM}
In the absence of PFDM the line element for the Hayward black hole can be written as
\cite{Ma:2020dhv}
\begin{eqnarray}
ds^{2} &=&-f(r)dt^2+\frac{1}{f(r)}dr^2+r^2(d\theta^2+\sin^2\theta d\phi^2),
\end{eqnarray}
where
\begin{eqnarray}
f(r)=1-\frac{2 m r^2}{r^3+Q_m^3}.
\end{eqnarray}%
The generalized off-shell free energy, the radial component of $n$ vector, and its zero point of the vector field in this case take the form
\begin{eqnarray}
\mathcal{F}&=&\frac{1}{2r^2_h\tau}\left\{-2\pi r^4_h+\tau r^3_h+Q^3\left(\tau-4\pi r_h\right) \right\},\\
\phi^{r_h}&=&\frac{1}{2r^3_h\tau}\left\{\tau \left(r^3_h-2Q_m^3\right)-4\pi r_h\left(r^3_h+Q^3_m\right)\right\},\\
\tau &=&\frac{4\pi r_h\left(r^3_h+Q^3\right)}{r^3_h-2Q^3}.
\end{eqnarray}
To find the winding numbers the complex function is given as
\begin{eqnarray}
    \mathcal{R}_s(z)=-\frac{z^3-2Q^3}{4\pi z^4-\tau z^3+4\pi Q^3z+2\tau Q^3}\equiv-\frac{z^3-2Q^3}{\mathcal{A}(z)}.
\end{eqnarray}
The root analysis of $\mathcal{A}(z)$ shows that for every $Q$ and $\tau_c<\tau$, $\mathcal{A}(z)$ has two real positive real root\footnote{The discriminant of $\mathcal{A}(z)$ is $\delta=-108\left(1024\pi^6 Q^{12}-1664\pi^3\tau^3 Q^3+\tau^6 Q^6\right)$ which is negative for all $Q$ and $\tau_c<\tau$.}. 
Thus, the winding and hence topological numbers for the Hayward black hole in the absence of dark matter are
\begin{eqnarray}
    w_1=-1,~ w_2=1 ~\text{and}~ W=0,
\end{eqnarray}
which are different from that of the Schwarschild black hole. This shows that the presence of the magnetic charge $Q_m$ in Hayward spacetime really affects both winding and topological numbers. This can be verified from the $n$ vector field [see Fig. \ref{vfHBHs} (top left panel)].

\begin{table} [b] 
\centering
\caption{In Table I, we  summarize our results for the  topological numbers as well as the generation point (GP) and annihilation point (AP) of the black holes studied in the present paper. We also present some of the results reported in the literature.}

\label{}
\begin{tabular}{|l|l|l|l|l|l|l|} 
\hline
Black hole spacetime & $W$ & GP & AP  \\
 \hline
Schwarzschild BH \cite{Wei:2022dzw} & -1 & 0 & 0  \\
Schwarzschild-AdS$_4$ BH  \cite{Yerra} & 0 & 0 & 1 \\ 
Kerr BH \cite{Wu:2022whe} & 0 & $1$ & $0$  \\ 
Kerr-AdS$_4$ BH  \cite{Wu:2022whe} & 1 & $1$ or $0$ & $1$ or $0$\\ 
Reissner-Nordstr$\Ddot{o}$m BH \cite{Wei:2022dzw} & 0 & $1$ & $0$  \\ 
Reissner-Nordstr$\Ddot{o}$m-AdS$_4$ BH \cite{Wei:2022dzw} & 1 & $1$ or $0$ & $1$ or $0$  \\ 
Kerr-Newman BH \cite{Wu:2022whe} & 0 & $1$ & $0$  \\ 
Kerr-Newman-AdS$_4$ BH \cite{Wu:2022whe} & 1 & $0$ & $0$  \\ 
Kerr BH in PFDM (present paper) & 0 & $1$ & $0$   \\ 
Schwarzschild BH in PFDM (present paper)  & -1 & $0$ & $0$  \\ 
Kerr AdS BH in PFDM (present paper) & 1 & $1$ or $0$ & $1$ or $0$   \\
Kerr-Newman BH in PFDM (present paper) & 0 & $1$ & $0$  \\
Reissner-Nordstr$\Ddot{o}$m BH in PFDM (present paper) & 0 & $1$ & $0$   \\
Hayward BH (present paper) & 0 & $1$ & $0$  \\
Rotating Hayward BH (present paper) & 0 & $1$ & $0$  \\
Static Hayward BH in PFDM (present paper) & 1 & $1$ or $0$ & $1$ or $0$  \\
Rotating Hayward BH in PFDM (present paper) & 0 & $1$ & $0$\\
\hline
\end{tabular}
\end{table}

\subsection{Static solution: Hayward BH in PFDM}
It has the line element \cite{Ma:2020dhv}
\begin{eqnarray}
ds^{2} &=&-f(r)dt^2+\frac{1}{f(r)}dr^2+r^2(d\theta^2+\sin^2\theta d\phi^2),
\end{eqnarray}%
with
\begin{eqnarray}
f(r)=1-\frac{2 m r^2}{r^3+Q_m^3}+\frac{\alpha}{r} \ln\left( \frac{r}{|\alpha |}\right),
\end{eqnarray}%
where $Q_m$ is the magnetic monopole charge. The mass $M$, magnetic charge $Q_m$, entropy $S$, and the Hawking temperature $T$ of the black hole are
\begin{eqnarray}
M&=&m,\\
S&=&=\pi \left[r_h^2-\frac{2 Q_m^3}{r_h}\right] \\
T&=& \frac{f'(r)}{4 \pi}|_{r_h}
\end{eqnarray}
where 
\begin{equation}
   f(r)=1-\frac{2 m r^2}{r^3+Q_m^3}+\frac{\alpha}{r} \ln\left( \frac{r}{|\alpha |}\right).
\end{equation}
The generalized off-shell free shell energy, the radial components of $n$ vector field, and its zero point can be obtained as
\begin{eqnarray}\notag
\mathcal{F}&=&\frac{1}{2r^3_h\tau}\left\{\left(r^3_h+Q^3\right)\left(r_h+\alpha\ln\left(\frac{r_h}{|\alpha|}\right)\right)-\pi\left(r^3_h-2Q^3\right)\right\},\\\notag
\phi^{r_h}&=&\frac{1}{2r^4_h\tau}\Big[r^3_h\Big\{\tau\left(r^2_h+\alpha r_h\right)-4\pi r^3_h\Big\}\\\notag
&-&Q^2_m\tau \Big\{2r^2_h+\alpha r_h\Big(1-3\ln\left(\frac{r_h}{|\alpha|}\right)+\frac{4\pi r^3_h}{\tau}\Big)\Big\}\Big],\\
\tau &=& 4\pi\frac{r^3_h\left(r^3_h+Q^3_m \right) }{r^3_h\left(r^2_h+\alpha r_h\right)-Q^3_m\left(2r^2_h-\alpha r_h+3\alpha r_h \ln\Big(\frac{r_h}{|\alpha |}\Big)\right)}.
\end{eqnarray}
The complex function in this case takes the form
\begin{eqnarray}
    \mathcal{R}(z)\equiv-\frac{z^3\left(z^2+\alpha z\right)-Q_m^3\Big(2z^2-\alpha z+3\alpha z \ln\Big(\frac{z}{|\alpha|}\Big)\Big)}{\mathcal{A}(z)},
\end{eqnarray}
where
\begin{eqnarray}\label{azfnRHBH}\notag
\mathcal{A}(z)&=&4\pi z^3\left(z^3+Q^3_m\right)-\tau\Big\{z^3\left(z^2+\alpha z\right)\\&-&Q_m^3 \Big(2z^2-\alpha z+3\alpha z \ln\Big(\frac{z}{|\alpha|}\Big)\Big)\Big\}. 
\end{eqnarray}
For the winding and topological number we have plotted the function $\mathcal{A}(z)$ in Fig. \ref{AzfHBHDMB}, which shows that unlike to Hayward black hole and rotating Hayward black hole in PFDM, the Hayward black hole in PFDM, for $\tau_c<\tau$, the equation $\mathcal{A}(z)$ has three real positive roots. So, similar to other cases we can easily deduce that the winding and the topological numbers for the Hayward black hole in PDFM are
\begin{eqnarray}
    w_1=-1, ~w_2=1, ~w_3=1 \quad \text{and}\quad W=1.
\end{eqnarray}
This case be verified from the vector field $\phi$ graph plotted in Fig. \ref{vfHBHs} (bottom right panel). Further, the zero point of $\phi$ is plotted in Fig. \ref{zeropHBHDM} which shows that the spacetime has generation as well as annihilation point. 

\subsection{Rotating Hayward BH}

Let us now elaborate the final special case, namely, we include the role of rotation and magnetic charge. For this case we get 
\begin{eqnarray}
\tau=4\pi\frac{r^6_h+Q_m^3r_h\left(r^2_h+a^2\right)}{r^3_h\left(r^2_h-a^2\right)-Q_m^3\left(4a^2+2r^2_h\right)}.
\end{eqnarray}

The function $\mathcal{R}(z)$ on the other hand takes the following form
\begin{eqnarray}
    \mathcal{R}(z)\equiv-\frac{z^3\left(z^2-a^2\right)-Q_m^3\left(4a^2+2z^2\right)}{\mathcal{A}(z)}
\end{eqnarray}
where we have defined
\begin{eqnarray}\label{azRHBH}
\mathcal{A}(z)&=& 4\pi\left\{z^6+Q^3_mz\left(z^2+a^2\right)\right\}\\\nonumber
&-&\tau\left\{z^3\left(z^2-a^2\right)-Q_m^3\left(4a^2+2z^2\right)\right\} 
\end{eqnarray}

The graph of the function $\mathcal{A}(z)$ is plotted in Fig. \ref{AzfRHBHDMB} which shows that for $\tau_c<\tau$, $\mathcal{A}(z)=0$ has two positive real roots and hence similar to that of the Hayward black hole, the rotating Hayward black hole has the same winding and topological numbers 
\begin{eqnarray}
    w_1=-1,~ w_2=1 ~\text{and}~ W=0.
\end{eqnarray}

\section{Conclusions}
We have extensively investigated the thermodynamic classes of a specific class of black hole solutions (singular and regular solutions) situated within the context of PFDM spacetime, all while accounting for rotational dynamics, electric charge, and magnetic charge effects. From a physical point of view, such a perfect fluid matter is of particular interest since it mimics the behavior of dark matter in large distances from the black hole.  The first important result of our analysis unveils that the winding and topological properties exhibited by Schwarzschild and Kerr black holes remain unaltered when the effect of PFDM is introduced. However, when considering the Kerr-AdS background in the realm of PFDM, a distinct topological feature emerges, setting it apart from the Kerr black hole in PFDM. Although, the same topological number is obtained compared to the Kerr-AdS black holes.

In the subsequent sections of this paper we have investigated the impact of electric charge and nonlinear magnetic charge on the thermodynamic topological classes. This exploration is conducted using both the Kerr-Newman black hole and the regular rotating Hayward black holes in PFDM. To this end, we analysed also the static metrics such as the Reissner-Nordstr$\Ddot{o}$m BH and the Hayward black holes in PFDM. It is very interesting to note that, if we compare our results with the existing results of the black holes in the literature, our findings demonstrate that there is no effect of PFDM on topological numbers for Schwarzschild BH, Kerr BH, Kerr-AdS BH, Reissner-Nordstr$\Ddot{o}$m BH and Kerr-Newman BH.  

In the final part of the present paper, we have studied the most interesting case that has to do with the role of the magnetic charges which can shift the topological numbers, accentuating differences between static Hayward BH and the Schwarzschild black hole. This difference in topological number compared to the Schwarzschild black hole arises in both cases that includes the absence and presence of PFDM. Moreover, when a magnetic charge is introduced, rotation exhibits a notable influence on the topological attributes as well. Table I provides a concise overview of our findings regarding the topological numbers, along with the generation point (GP) and annihilation point (AP) for the black holes investigated in this study. Additionally, we incorporate some of the most important cases studied in existing literature for comparative context.

\end{document}